\documentclass{article}
\usepackage{graphicx} 
\usepackage{authblk}
\newcommand{\keywords}[1]{\par\noindent\textbf{Keywords:} #1}

\title{DepoRanker: A Web Tool to predict \textit{Klebsiella} Depolymerases using Machine Learning}

\author[a,1]{George Wright}
\author[b]{Slawomir Michniewski}
\author[c]{Eleanor Jameson}
\author[a,2]{Fayyaz ul Amir Afsar Minhas}

\affil[a]{Department of Computer Science, University of Warwick, 6 Lord Bhattacharyya Way, Coventry, CV4 7EZ, UK.}
\affil[b]{Department of Genetics and Genome Biology, University of Leicester, Leicester, LE1 7RH, UK}
\affil[c]{School of Environmental and Natural Sciences, Bangor University, Bangor, LL57 2DG, UK.}

\begin{document}

\maketitle


\begin{abstract}


Background: Phage therapy shows promise for treating antibiotic-resistant \textit{Klebsiella} infections. Identifying phage depolymerases that target \textit{Klebsiella} capsular polysaccharides is crucial, as these capsules contribute to biofilm formation and virulence. However, homology-based searches have limitations in novel depolymerase discovery. 

Objective: To develop a machine learning model for identifying and ranking potential phage depolymerases targeting \textit{Klebsiella}.

Methods: We developed DepoRanker, a machine learning algorithm to rank proteins by their likelihood of being depolymerases. The model was experimentally validated on 5 newly characterized proteins and compared to BLAST. 

Results: DepoRanker demonstrated superior performance to BLAST in identifying potential depolymerases. Experimental validation confirmed its predictive ability on novel proteins. 

Conclusions: DepoRanker provides an accurate and functional tool to expedite depolymerase discovery for phage therapy against \textit{Klebsiella}. It is available as a webserver and open-source software. 

Availability: Webserver: https://deporanker.dcs.warwick.ac.uk/
Source code: https://github.com/wgrgwrght/deporanker

\end{abstract}

\section*{Introduction}

Antimicrobial resistance (AMR) stands as a critical global health threat, with the emergence of multidrug-resistant Enterobacteriaceae infections presenting a serious challenge to the efficacy of current antibiotics in clinical settings~\cite{Tacconelli2017}. The ability of the Enterobacteriaceae to gain and transfer an increasing number of antimicrobial-resistance genes, particularly in health-care settings represents a critical threat to human health~\cite{navon-venezia_klebsiella_2017}. \textit{Klebsiella}, a prominent member of the Enterobacteriaceae family, has been designated by the World Health Organization as a priority pathogen, demanding urgent efforts for the development of novel antimicrobials due to its heightened resistance levels~\cite{TACCONELLI2018318}. In the face of this growing crisis, phage therapy is being explored as a potential solution to combat AMR infections. This approach harnesses bacteriophages, also referred to as phages, to selectively target and eliminate bacterial pathogens~\cite{dmphage}.  

\textit{Klebsiella} phages exhibit a remarkably narrow host range, further constrained by the diverse polysaccharide capsules expressed by their bacterial hosts. These capsules serve as protective barriers, rendering the host bacteria inherently resistant to both antibiotic treatment and phage infections~\cite{kochan2023klebsiella}. Importantly, these capsules contribute to the problematic nature of \textit{Klebsiella} infections and are been linked to biofilm formation and virulence~\cite{li2022characterization}. Notably, certain \textit{Klebsiella} phages display specificity towards particular host capsule types, a phenomenon often associated with the presence of sugar-degrading enzymes known as depolymerases~\cite{solovieva_comparative_2018,hughes_biofilm_1998}. Phage depolymerases have the added value of being applied as therapeutic enzymes in their own right which circumvents some of the obstacles to using phages as antimicrobials~\cite{wang2023translating}. A limited set of depolymerases have previously been characterised through traditional wet-lab studies, highlighting the time-intensive nature of this process~\cite{ hsu_isolation_2013, majkowska-skrobek_capsule-targeting_2016,pan_klebsiella_2017, 10.3389/fmicb.2019.02949}.

Given the challenges associated with laboratory based characterization, the integration of computational tools becomes crucial in expediting the detection of depolymerases within bacteriophage proteomes. Computational approaches hold the potential to streamline and enhance the identification of depolymerases, facilitating the development and deployment of phages for therapeutic purposes. Traditional methodologies for targeted detection of \textit{Klebsiella} phage depolymerases have relied on homology searches such as BLAST~\cite{Tatusova1999-mw}. This approach aides the refinement of searches by identifying “tail-fibre” or “tail-spike” proteins~\cite{squeglia_structural_2020, latka_modelling_2019}. However, some phages without recognisable depolymerase genes display depolymerase-like activity in laboratory experiments. In these cases, the effectiveness of sequence-based annotation for phage genomes could be compromised, since they rely on the similarity to existing depolymerases~\cite{mcnair_phage_2018,mcnair_phanotate_2019}.

In this paper, we introduce DepoRanker, the first machine learning algorithm designed to enhance traditional bioinformatics pipelines for depolymerase identification~\cite{wrightmachine}. DepoRanker surpasses the performance of BLAST and demonstrates robust generalization capabilities to an independent test set comprising five phage proteomes with recently characterized depolymerases. Additionally, to enhance the accessibility and applicability of the model, we have developed a user-friendly web server. This platform enables seamless integration of our model into clinical workflows. Through the implementation of this model and web server, our aim is to enhance the precision and efficiency of depolymerase discovery.

\section*{Methods and Materials}

\subsection*{Data Collection}

We collected data from previously published sources containing experimentally verified depolymerase proteins isolated from phages~\cite{DUNSTAN2023112551, thung2021component, li2022identification, li2022characterization}. Within this dataset, all depolymerases were classified as the positive class. To construct the negative dataset, we leveraged the phage proteomes associated with these proteins. Following the extraction of all proteins and the removal of those already classified as depolymerases, we formed the negative dataset. This process yielded a final dataset comprising 24 phage proteomes, encompassing a total of 39 characterized depolymerase protein sequences and 2,601 non-depolymerase protein sequences. For a comprehensive list of the accession numbers corresponding to all proteins utilized, please refer to Table(~\ref{table:res1}).

In our evaluation of the predictive accuracy of DepoRanker, we also assembled an external dataset comprising five recently characterized proteins known for their depolymerase activity, along with the associated phage proteomes. The list of proteins with accession numbers can be found in table(~\ref{table:res2}) along with proteome size. This dataset was utilized as an external test set to assess the model's ability to generalize.

\subsection*{Feature Extraction}

Protein sequences are given in FASTA format, the sequence data was then extracted from this datatype and converted into a string of letters taken from an alphabet of length 20 representing the 20 different amino acids. In this work, a simple amino acid composition of a protein is used as its feature representation~\cite{biology9110365,9198006}. The count of each letter in the sequence is taken giving a 20-dimensional non-normalised feature vector representation for a given protein sequence.

\subsection*{Machine Learning Model}

In this paper, we present a novel implementation of Extreme Gradient Boosting (XGBoost)~\cite{ChenG16} for ranking proteins in a pairwise manner. The XGBoost ranking model is trained on multiple training proteomes to assign high scores to known depolymerases and low scores to non-depolymerases within each input phage. During testing, this approach is replicated to generate a ranked list of proteins in the test phage, prioritizing them based on predicted depolymerase activity. Our objective is to obtain a prediction function from XGBoost, $f(x, \theta)$, with learnable parameters $\theta$ for a protein sequence represented in terms of its feature vector $x$. We require the model to learn the optimal parameters $\theta^*$ such that the score $f(x_i, \theta^*)$ for positive examples $x_i \in P_g$ (known depolymerases in phage $g$) is higher than $f(x_j, \theta^*)$ for $x_j \in N_g$ (non-depolymerase from the same phage) across all training phages. The hyper parameters of the learning model are selected through the cross-validation performance evaluation and the optimal results obtained are a learning rate of 0.1, a subsample of 0.9 and max tree depth of 3. The XGBoost model is trained on all proteins in the training set to produce a model for performance evaluation. The same model has previously been utilized for the discovery of novel anti-CRISPR proteins~\cite{10.1093/nar/gkaa219}. 

The model was trained in a pairwise fashion such that each depolymerase had a higher score than non-depolymerase proteins. In testing, the same approach is used for generation of a ranked list of proteins in the test phage based on predicted depolymerase activity.  The XGBoost model is trained on all proteins in the training set to produce a model for performance evaluation. The prediction scores were then computed with this model on a proteome-by-proteome basis to investigate whether known depolymerases scored highly and were among the top in the ranked list.

\subsection*{Non-redundant Cross-Validation Folds}

To ensure rigorous testing, we utilized CD-HIT~\cite{Huang2010} to  group the 39 identified depolymerases based on sequence similarity to
generate non-redundant sets. These sets were then mapped to the respective phage proteomes associated with each depolymerase. If a phage contained multiple depolymerases separated into different clusters, then this phage would appear in multiple training sets. To ensure the non-redundancy of our training sets, in this scenario the sets were merged. Depolymerases were clustered using a sequence threshold of 10\%; any depolymerase exhibiting a similarity score exceeding 10\% was deemed to potentially possess the same structure and function. This gave 7 different clusters. This approach ensures that the model is not trained on sequences similar to those in the test set, thereby enhancing its robustness. To assess the efficacy of our machine learning model, we conducted non-redundant cross-validation on the data by employing these 7 non-redundant training sets obtained by clustering the proteins from CD-HIT as the folds for cross-validation. Proteomes allocated to the fold were reserved for validation, while the remaining proteomes constituted the training set. This resulted in 7 different models from the cross-validation process. For a final model, an ensemble of these 7 models is used. An instance is passed through each model and the mean score is taken for a final output score.

\subsection*{Performance Evaluation}

To assess the predictive performance of DepoRanker, we use several metrics including the Area Under the Receiver Operating Characteristic Curve (AUROC), the Area Under the Precision-Recall Curve (AUCPR) and the rank of the first positive prediction (RFPP)~\cite{Minhas2014-oc}. RFPP reports the rank of the top scoring depolymerase, emphasizing that depolymerases should receive higher scores and consequently rank lower in a sorted list than non-depolymerase proteins. This approach aligns with the practical goal of discovering depolymerases with minimal wet lab experiments. Intuitively, this means a perfect prediction model would yield an RFPP of 1 for all proteomes, indicating that only one laboratory experiment would be necessary to identify a depolymerase in the phage proteome. Using AUROC and AUCPR further allows us to evaluate the model's overall performance in distinguishing between depolymerases and non-depolymerase proteins across different thresholds, providing a comprehensive assessment of its predictive capabilities.

\subsection*{Comparison with homology searches}

Since no models exist which use machine learning to solve this task, we have used BLAST as a baseline for comparison~\cite{Tatusova1999-mw}. BLAST is a widely employed tcool for sequence alignment which searches for similar sequences in a database based on sequence homology, and assigns a score to each match. To ensure a fair evaluation, BLAST was run on the same dataset used for training the machine learning model. All test proteins in a phage were searched against the dataset of known depolymerases and sorted based on their e-values to rank proteins based on their expected depolymerase activity.

\section*{Results}

\subsection*{Cross Validation Results}

Our model has a median RFPP of 1, meaning that for each proteome tested during cross validation 50\% would have a known depolymerase in as the top prediction by the model. For the 100th percentile our model shows an RFPP of 3. Meaning out of the 24 phage proteomes tested, the ranking model had an RFPP of 3 or lower for all proteomes, which is in the top 5\% in their respective proteome. This is compared to a random ranking model that produces a random score for any given example, which gives a median RFPP is 52, and BLAST which has a RFPP of 31 for the 100th percentile. Fig. 1 shows the RFPP across percentiles of the results for all proteomes across different models. This clearly shows the effectiveness of our ranking model for identifying depolymerases in phage proteomes. For the proteomes with an RFPP of 2 and 3 there are two possibilities for these non-depolymerase proteins being ranked higher than known depolymerases; 1. they are false positives, 2. they are true depolymerases that were not characterised, present in the proteomes of phages that also contained a characterised depolymerase. Table 1 shows the full cross validation results showing BLAST fails for some proteomes as a result of relying on sequence similarity in the training data.  This underscores that our machine learning model captures more nuanced information about depolymerases than a simplistic sequence comparison, making it a more reliable tool for depolymerase discovery. 

To further evaluate the performance of our ranking model, we calculated the AUROC and the AUCPR, as shown in Figure~\ref{fig:ROC}. Our model achieved an AUROC of 0.99, indicating excellent discrimination between depolymerases and non-depolymerases. The AUCPR was 0.42, demonstrating a high level of precision and recall in identifying true depolymerases. These metrics highlight the model's ability to accurately rank depolymerases at the top while minimizing false positives. In comparison, the AUROC and AUCPR for BLAST were 0.94 and 0.37, respectively, further underscoring the superiority of our machine learning approach over traditional sequence similarity methods.

\subsection*{External Test Results}

Additionally, an external test set comprising of 5 recently characterized proteins and phage proteomes assessed the model's generalization capabilities. Table 2 shows the predictions generated by our model revealing a good performance in discriminating proteins with depolymerase activity showing that the model generalises to external proteomes. For 3 proteins, DepoRanker scores perfectly, and for all 5 proteins DepoRanker scores a depolymerase in the top 3 which corresponds with the cross validation results. This performance proves the models capacity to extend predictions to proteins beyond the training set, proving overall utility and effectiveness in a real-world application.

\subsection*{Feature Importance}

To assess the significance of amino acid count features, we utilized SHAP values which elucidates feature importance by attributing contributions to individual features for each prediction. The mean absolute SHAP value across folds was computed to assess the impact of each feature on the model output. Our analysis unveiled that three of the top four amino acids influencing the model's predictions possess hydrophobic side chains, namely Tryptophan (W), Alanine (A), and Tyrosine (Y). This observation underscores the importance of hydrophobic amino acids in protein structure and function in depolymerases. The top 5 features with the highest mean absolute SHAP values from non-redundant cross-validation are illustrated in Fig. \ref{fig:SHAP}, providing a visual representation of their impact.

\subsection*{Availability}

To increase the accessibility and allow this to be used by any researcher working on identifying depolymerases, a web server has been built and is accessible through a standard browser at https://deporanker.dcs.warwick.ac.uk/, with the source code for the model and website is also available at the github repository https://github.com/wgrgwrght/DepoRanker. The input for DepoRanker is a list of complete protein sequences in FASTA format. For clarity, the website includes an illustrative example file. While the proteomes need not be exclusively from \textit{Klebsiella}, it is essential to note that the model's training primarily encompasses this bacterial genus. Therefore, the generalizability of the model to phages from other species is not clearly established. The functionality employed by the website is implemented through a HTML form. On pressing the rank proteome button, a HTML page is returned containing a link to a downloadable CSV file for subsequent offline analysis. The CSV file contains the protein ID, rank, and model output score, with all proteins ranked in descending order by the models output score. A protein with a higher score is predicted to be more likely to act as a depolymerase than a protein with a lower score.  

We have processed 665 \textit{Klebsiella} phage proteomes using the DepoRanker model and include the highest scoring protein for each in a list. The sequence information for each of these proteins is available for download in FASTA format along with the, protein ID, associated phage accession number, and output score. Since the output scores of the model may be seen as a confidence value, if the scores for all proteins are low, then this might suggest a phage which does not have depolymerase behaviour. This compilation serves as a valuable resource, enabling thorough examination and exploration of depolymerases for future research initiatives.

\section*{Discussion}

In response to the escalating threat of AMR \textit{Klebsiella} infections, we introduce DepoRanker, the first machine learning model designed to expedite the identification of depolymerases within phage proteomes. By leveraging advanced algorithms, DepoRanker outperforms traditional homology-based searches like BLAST, achieving superior accuracy in ranking proteins based on their potential as depolymerases. Specifically, on non-redundant folds of the training data, DepoRanker demonstrated a maximum RFPP of 3, compared to BLAST's maximum RFPP of 31 indicating superior ranking capability. Additionally to assess generalization, we evaluated DepoRanker on 5 external test proteomes not included in the training set. The model exhibited strong performance, with an average RFPP of 1.6, demonstrating its ability to generalize beyond the specific proteins used for training.

To enhance accessibility, we have implemented DepoRanker as a user-friendly web server, enabling researchers without programming expertise to harness the benefits of this powerful model. By making advanced machine learning techniques available, the depolymerase identification process is greatly simplified and streamlined, enabling phage researchers to leverage the power of cutting-edge computational methods to efficiently assess the therapeutic potential of phages against AMR \textit{Klebsiella}  infections. Additionally, we provide an extensive list of putative depolymerases identified by our model, fostering reuse and further exploration within the research community. 

As a next step, any novel depolymerase groups identified through our machine learning approach will require rigorous structural and biochemical validation to confirm their activity and therapeutic potential. These validation efforts will contribute to the broader understanding of phage therapeutics in the context of antibiotic resistance as well as confirming the accuracy of our predictions. While our results are promising, we recognize potential limitations and challenges associated with our approach. One limitation is the reliance on the quality and diversity of the training data, which may not capture the full spectrum of depolymerase diversity. To address this, the training set could be continuously updated as new depolymerase sequences become available, ensuring that DepoRanker remains robust and comprehensive. While DepoRanker was developed specifically for identifying depolymerases against Klebsiella, the underlying approach and architecture could potentially be adapted to identify other types of therapeutic proteins or target different pathogens. By leveraging the power of machine learning and deep neural networks, our methodology could be extended to accelerate the discovery and development of novel antimicrobial therapies against a wide range of AMR threats.

\section*{Conclusion}

Phage therapy represents a promising approach for combating AMR \textit{Klebsiella}   infections, and the identification of effective depolymerases is a crucial step in this endeavour. DepoRanker, our novel machine learning algorithm, addresses the limitations of traditional methods and significantly streamlines the discovery process. By providing a web server and open-source code, we ensure that DepoRanker is accessible to the broader scientific community, fostering further advancements in the field of phage therapy research. The integration of machine learning in this context not only enhances the efficiency and precision of the discovery process but also paves the way for future innovations in phage-based therapeutic strategies against antibiotic-resistant infections.

\section*{Acknowledgments}

The authors would like to express their sincere gratitude to the IT support staff at the University of Warwick Mohammed Bashar Habib and Salinder Tandi for their invaluable assistance in setting up the computing server used for this research.

\section*{Authorship contribution statement}
GW: Formal analysis(lead); Software(lead); Visualization(lead); Writing – original draft(equal).
SM: Data curation(equal); Validation(equal); Writing – original draft(equal); Writing – review \& editing(equal).
FM: Conceptualization(equal); Methodology(lead); Supervision(lead); Writing – original draft(equal); Writing – review \& editing(equal).
EJ: Conceptualization(equal); Data curation(equal); Validation(equal); Writing – original draft(equal); Writing – review \& editing(equal).

\section*{Authors disclosure statement}
The authors declare no conflicts of interest.

\section*{Funding statement}
No funding was received for this work.

\newpage

\section*{Tables}

\begin{table}[!ht]
\begin{center}
\caption{Results for non-redundant cross-validation. Each phage is referred by its accession number with the ranks of the top scoring depolymerase protein from different methods for each test phage shown together with the size of the proteome.}
\begin{tabular}{||c c c c||} 
\hline
Phage Accession Number & Proteome Size & BLAST & DepoRanker \\ \hline
NC027399 & 540 & 1 & 1 \\ \hline
AB797215 & 203 & 20 & 1 \\ \hline
MW655991 & 76 & 3 & 1 \\ \hline
MW672037 & 74 & 3 & 1 \\ \hline
MF663761 & 71 & 1 & 1 \\ \hline
OU509534 & 56 & 4 & 1 \\ \hline
OU509533 & 54 & 2 & 1 \\ \hline
AB716666 & 53 & 4 & 1 \\ \hline
KX712070 & 53 & 1 & 1 \\ \hline
KU666550 & 52 & 2 & 1 \\ \hline
KU183006 & 51 & 1 & 1 \\ \hline
GQ413937 & 49 & 1 & 1 \\ \hline
MK903728 & 49 & 1 & 1 \\ \hline
MT966872 & 48 & 1 & 1 \\ \hline
KY389315 & 48 & 1 & 1 \\ \hline
LC413194 & 47 & 1 & 1 \\ \hline
JF501022 & 77 & 31 & 2 \\ \hline
KY385423 & 54 & 2 & 2 \\ \hline
KT964103 & 54 & 2 & 2 \\ \hline
KY389316 & 47 & 1 & 2 \\ \hline
OU509535 & 528 & 8 & 3 \\ \hline
LC413195 & 53 & 1 & 3 \\ \hline
MH587638 & 50 & 1 & 3 \\ \hline
LC413193 & 48 & 1 & 3 \\ \hline
\end{tabular}
\label{table:res1}
\end{center}
\end{table}

\begin{table}[!ht]
\centering
\caption{Results on the external test. Each phage is referred by its accession number, with the rank of the top scoring depolymerase protein for each test phage shown with the size of the proteome.}
\begin{tabular}{||c c c||} 
\hline
{Phage Accession Number} & { Proteome Size} & { DepoRanker} \\ \hline
{OM938992}               & { 78}            & { 1}          \\ \hline
{ MT894004}               & { 67}            & { 1}          \\ \hline
{ OM938991}               & { 76}            & { 1}          \\ \hline
{ MT966873}               & { 47}            & { 2}          \\ \hline
{ MW722080}               & { 78}            & { 3}          \\ \hline
\end{tabular}
\label{table:res2}
\end{table}

\newpage

\section*{Figures}

\begin{figure}[!ht]
  \centering
  \includegraphics[width=1.0\textwidth]{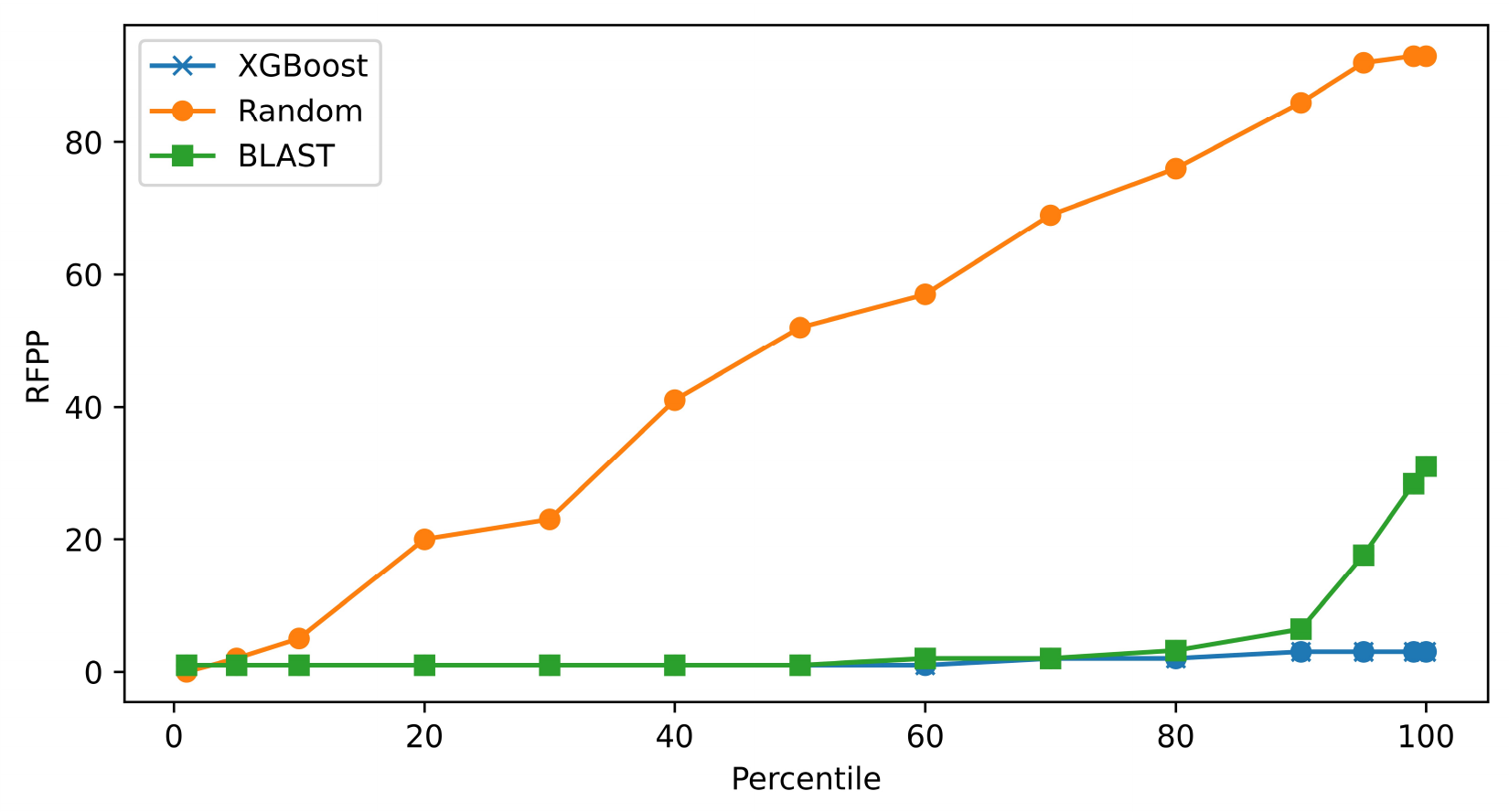}
  \caption{Comparison between different models using percentiles of Rank of First Positive Prediction (RFPP). The curves show RFPP percentiles for the proposed model, a random baseline, and BLAST.}
  \label{RFPP}
\end{figure}

\begin{figure}[!ht]
  \centering
  \includegraphics[width=1.0\textwidth]{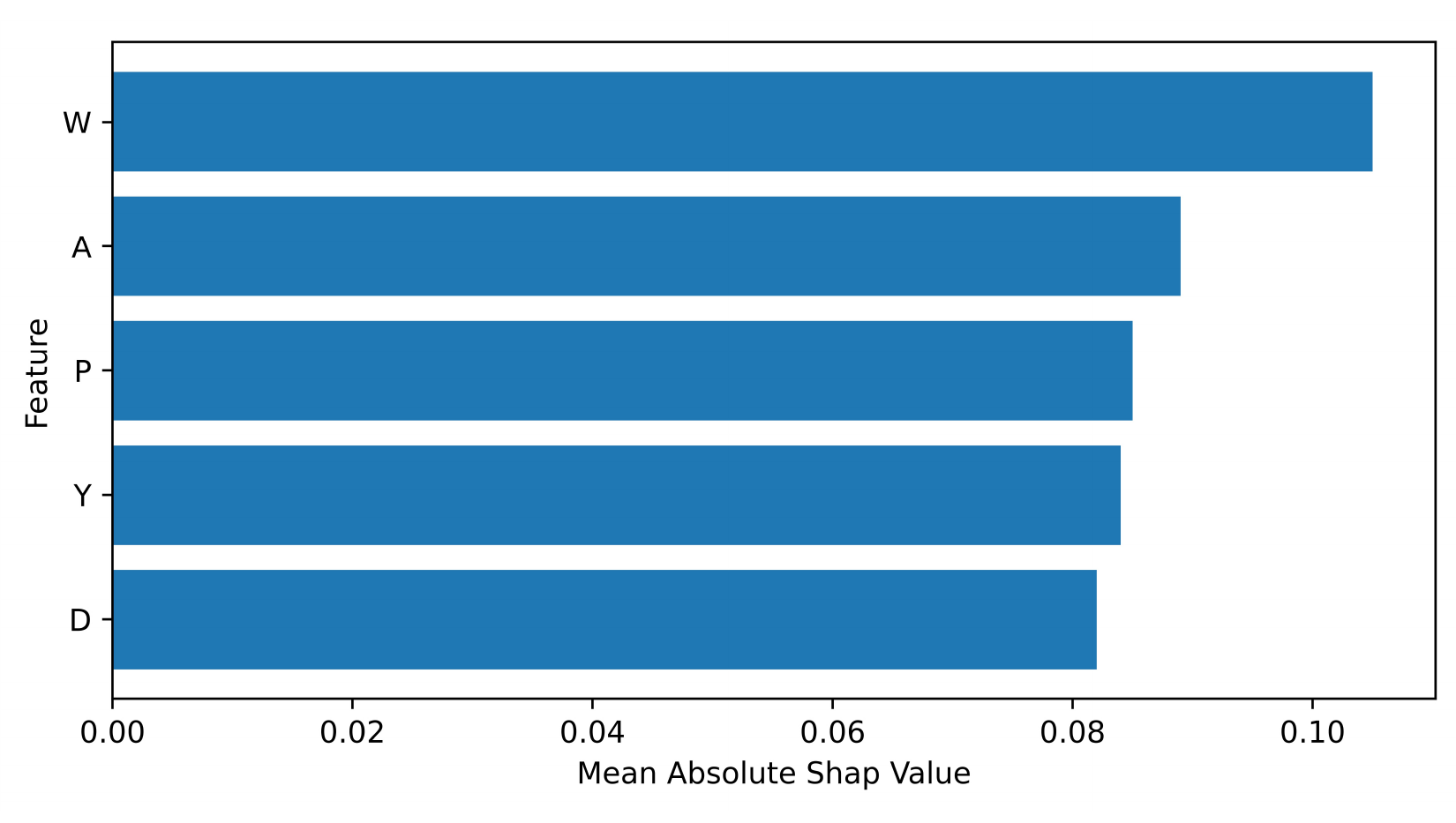}
  \caption{The top 5 mean absolute SHAP values for each feature for scores from cross-validation.}
  \label{fig:SHAP}
\end{figure}

\begin{figure}[!ht]
  \centering
  \includegraphics[width=1.0\textwidth]{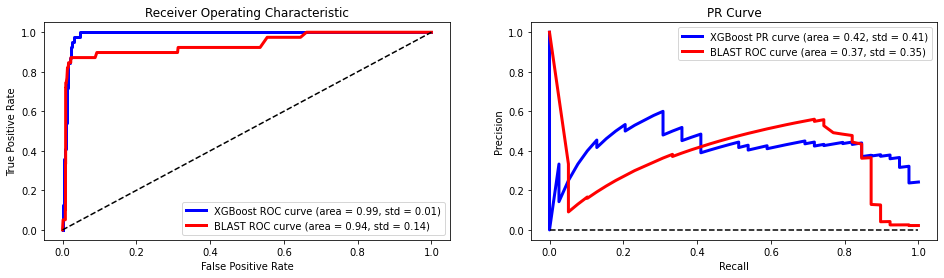}
  \caption{Receiver operating characteristic (ROC) curves and showing the performance for our ranking model (blue) and BLAST (red).}
  \label{fig:ROC}
\end{figure}

\end{document}